\documentclass[pra,twocolumn,showpacs,preprintnumbers,amsmath,amssymb,nofootinbib]{revtex4}

\usepackage{bm}

\newcommand{\upp}[1]{^{\rm \scriptscriptstyle #1}}
\newcommand{\dnn}[1]{_{\rm \scriptscriptstyle #1}}
\newcommand{\ket}[1]{|  #1 \rangle}
\newcommand{\bra}[1]{ \langle #1  |}
\newcommand{\proj}[1]{\ket{#1}\bra{#1}}

\newcommand{\abs}[1]{ | \, #1 \,  |}

\newcommand{\tr}{{\rm tr}}

\newcommand{\beq}{\begin{equation}}
\newcommand{\eeq}{\end{equation}}

\newcommand{\be}{\begin{equation}}
\newcommand{\ee}{\end{equation}}
\newcommand{\ba}{\begin{array}}
\newcommand{\ea}{\end{array}}

\newcommand{\asredchem}{\leadsto}

\newcommand{\storedchem}{\to}
\newcommand{\mulstoredchem}{\twoheadrightarrow}

\newtheorem{lem}{Lemma}
\newtheorem{theo}{Theorem}

\begin{document}

\title{The Power of LOCCq State Transformations}

\author{Ashish V. Thapliyal}
 \altaffiliation[Also at ]{Mathematical Sciences Research Institute, Berkeley, CA}
 \email{ash@msri.org, thaps@cs.berkeley.edu}
\affiliation{%
Department of Computer Science\\
University of California at Berkeley\\
Berkeley, CA 94720.
}%

\author{John A. Smolin}
\affiliation{IBM Thomas J. Watson Research Center\\ Yorktown
Heights, NY 10598.} \email{smolin@watson.ibm.com}
%

\date{\today}

\begin{abstract}
Reversible state transformations under entanglement non-increasing
operations give rise to entanglement measures.  It is well known
that asymptotic local operations and classical communication
(LOCC) are required to get a simple operational measure of
bipartite pure state entanglement.  For bipartite mixed states and
multipartite pure states it is likely that a more powerful class
of operations will be needed.  To this end \cite{BPRST:01} have
defined more powerful versions of state transformations (or
reducibilities), namely LOCCq (asymptotic LOCC with a sublinear
amount of quantum communication) and CLOCC (asymptotic LOCC with
catalysis).  In this paper we show that {\em LOCCq state
transformations are only as powerful as asymptotic LOCC state
transformations} for multipartite pure states.  We first
generalize the concept of entanglement gambling from two parties
to multiple parties: any pure multipartite entangled state can be
transformed to an EPR pair shared by some pair of parties and that
any irreducible $m$ $(m\ge 2)$ party pure state can be used to
create any other state (pure or mixed), using only local
operations and classical communication (LOCC).  We then use this
tool to prove the result.  We mention some applications of
multipartite entanglement gambling to multipartite distillability
and to characterizations of multipartite minimal entanglement
generating sets.  Finally we discuss generalizations of this
result to mixed states by defining the class of {\em cat
distillable states}.

\end{abstract}
\pacs{Valid PACS appear here}
\maketitle

\section{Introduction}
Entanglement is a fundamental aspect of quantum mechanics.  It has
been found useful for various information processing tasks such as
teleportation\cite{BBCJPW:93}, superdense coding\cite{BW:92},
entanglement assisted classical and quantum
communication\cite{BSST:99, BSST:02}, quantum
algorithms\cite{S:94}, and quantum cryptography\cite{BB:84}. Since
it is such an important resource, much effort has been put into
quantifying it. Entanglement for two-party pure states is
completely understood. For mixed states the situation is not as
clear, there being various different measures.   Recently,
\cite{BPRST:01} have proposed a program to quantify multipartite
entanglement using the idea of reversible state transformations
induced by entanglement non-increasing operations. It is well
known that asymptotic local operations and classical communication
(LOCC) are required to get a simple operational measure of
bipartite pure state entanglement. For bipartite mixed states and
multipartite pure states it is likely that a more powerful class
of operations will be needed. To this end \cite{BPRST:01} have
defined more powerful versions of state transformations (or
reducibilities), namely LOCCq (asymptotic LOCC with a sublinear
amount of quantum communication) and CLOCC (asymptotic LOCC with
catalysis).  In this paper we show that {\em LOCCq state
transformations are only as powerful as asymptotic LOCC state
transformations} for multipartite pure states.  We first
generalize the concept of entanglement gambling from two parties
to multiple parties: any pure multipartite entangled state can be
transformed to an EPR pair shared by some pair of parties and that
any non-trivial $m$ $(m\ge 2)$ party pure state can be used to
create any other state (pure or mixed), using only local
operations and classical communication (LOCC).  We then use this
tool to prove the result. We mention some applications of
multipartite entanglement gambling to multipartite distillability
and to characterizations of multipartite minimal entanglement
generating sets.  Finally we discuss generalizations of this
result to mixed states by defining the class of {\em cat
distillable states}.

\section{Multipartite Entanglement Gambling}
We begin by looking at entanglement gambling for bipartite pure
states.  Bennett, Bernstein, Popescu and Schumacher introduced the
idea of entanglement gambling in \cite{BBPS:96}. The idea is to
produce an EPR pair with a positive probability using local
operations and classical communication (LOCC) starting from any
other entangled bipartite pure state.  We briefly review the
bipartite entanglement portocol. Let us consider an arbitrary
entangled pure state $\Psi$ shared by A and B.  It is well known
that for a bipartite pure state can always be written in a Schmidt
decomposition
\begin{equation}
\ket{\Psi} = \sum_{i=1}^{k} a_i \ket{i\upp{A}i\upp{B}},
\end{equation}
where $k\ge 2$ since the state is entangled, $\{a_i>0 | i = 1,...,
n\}$, $\ket{i\upp{A}}$ form an orthonormal basis for $A$ and
$\ket{i\upp{B}}$ form an orthonormal basis for $B$.  Now A and B
can apply the local projectors $ P\upp{A/B} = \proj{0\upp{A/B}} +
\proj{1\upp{A/B}}$ on their halves of the state. This produces
state
$$\psi_1 =  c \ket{00} + d
\ket{11}$$ with probability $p = a_1^2 + a_2^2$, where $$c =
\frac{a_1}{p} \ \mathrm{and}\  d = \frac{a_2}{p}.$$ Then Alice
applies the local quantum operation given by the superoperator
with operator elements \begin{eqnarray*} & &A_1 = d \proj{0} + c\proj{1} \\
& &A_2 = \sqrt{1-d^2}\proj{0} + \sqrt{1-c^2} \proj{1},
\end{eqnarray*}
then the outcome corresponding to $A_1$ gives an EPR pair with
probability $2c^2d^2$.  Thus the total success probability for the
whole process is $(2 a_1^2 a_2^2)/(a_1^2 + a_2^2)$ which is
non-zero. Thus any pure bipartite entangled state can be converted
to an EPR pair with a positive probability.

Let us now write the above result in the notation used by
\cite{BPRST:01}\footnote{In \cite{BPRST:01} state transformations
are also called as reducibilities: If $\psi$ is transformed to
$\phi$ we can say that the problem of creating $\phi$ is reducible
to the problem of creating $\psi$. This provides the intuition
behind the name reducibility. In this paper we will use the state
transformations language instead of reducibilities.}.  First we
briefly review the notation.  We start with state transformations
for one copy of a state involving probabilistic outcomes, where
the procedure for the reducibility may fail some of time.  This is
called stochastic state transformation.

We say a state $\Psi$ is {\em stochastic LOCC transformable} to
$\Phi$ with yield $p$, written as $\Psi \storedchem\dnn{LOCC}
\Phi^{\otimes p}$ if and only if
\begin{equation}
\exists{{\mathcal L}}\;\;\; \Phi = \frac{{\mathcal L}(\Psi)}{\tr{}
{{\mathcal L} (\Psi)}} \enspace,
\end{equation}
where ${\mathcal L}$ is a multilocally implementable
superoperator\footnote{A multilocally implementable superoperator
is just a mathematical representation of a LOCC protocol.} such
that $\tr{}{{\mathcal L}(\Psi)} = p$. This means that a copy of
$\Phi$ may be obtained from a copy of $\Psi$ with probability $p$
by LOCC operations.  When $p=1$ it is called an exact
transformation or an exact reducibility.

Let $\mathcal{E}_2$ denote the set of bipartite pure entangled
states, then the bipartite entanglement gambling result can be
expressed as
\begin{equation}
\label{eq:gamble}
\forall \psi \in \mathcal{E}_2, \ \ \ \exists
p>0, \ \ \ \psi \storedchem \mathrm{EPR}^{\otimes p} .
\end{equation}

Clearly a generalized version of stochastic transformations is
obtained if we allow a finite number of copies of the source and
target states.   We say state $\Psi$ is {\em multicopy stochastic
LOCC transformable} to state $\Phi$ with yield $p$, written as
$\Psi \mulstoredchem\dnn{LOCC} \Phi^{\otimes p}$, if and only if
\begin{equation}
\label{eq:gengamble}
\exists_{{\mathcal L},m,n}\;\;\;
\Phi^{\otimes n} = \frac{{\mathcal L}(\Psi^{\otimes m})}{\tr{}
{{\mathcal L} (\Psi^{\otimes m})}} \enspace,
\end{equation}
where ${\mathcal L}$ is a multilocally implementable superoperator
such that $\tr{}{{\mathcal L}(\Psi)} = p m/n$. This means that $n$
copies of $\Phi$ may be obtained from $m$ copies of $\Psi$ with
yield $p$ per copy by LOCC operations.

Let us return to bipartite entanglement gambling again.  It gives
us an EPR pair with positive probability starting from any
entangled pure state. Since EPR pairs can be used in a
teleportation protocol to create an arbitrary bipartite state,
clearly any bipartite pure entangled state may be converted to any
other bipartite state with a positive probability.  Notice that
this protocol will in general require multiple copies of the
source state since the target state may be a state with higher
Schmidt number. Thus a stronger version of bipartite gambling can
be written using the multicopy stochastic reducibility as
\begin{equation}
\forall \psi \in \mathcal{E}_2, \ \ \ \exists p>0,   \ \ \  \psi
\mulstoredchem \phi^{\otimes p} ,
\end{equation}
where $\mathcal{E}_2$ denotes the set of bipartite pure entangled
states and $\phi$ is any bipartite state, pure or mixed, in finite
dimensions.

Now let us consider the multi-party scenario: There are  $m$
parties $(m\ge 2)$ labelled as $\{1, 2, ..., m\}$.  Given a
non-trivial subset $X$ of the parties and its complement
$\bar{X}$, we say that $\{X, \bar{X}\}$ defines a {\em cut}
between $X$ and $\bar{X}$.  We say that pure state $\Psi$ is {\em
factorizable} across the cut $ \{ X, \bar{X} \}$ of the parties if
$\Psi$ can be written  as a tensor product of two states, one with
the parties in set $X$ and the other with the parties in the
complement $\bar{X}$.  We say that a state is {\em entangled} if
it is not factorizable across some cut.  We define a pure state to
be {\em irreducible} if it is not factorizable across all cuts.
Thus an irreducible $m$-party pure state captures the notion of a
true $m$-party state.  Now we are ready to generalize entanglement
gambling.

It turns out that for multiple parties,  gambling can be
generalized in different ways.  First we generalize the weaker
result shown in equation \ref{eq:gamble}.  In this case we show
that an entangled pure multipartite state can be transformed under
LOCC to an EPR pair between some pair of parties.  We write this
as a lemma\footnote{This lemma was independently proved in
\cite{D:01}}.
\begin{lem}
\label{lem:someEPR} {\rm : If state $\Psi$ is an $m$-partite pure
state that is entangled  across the cut $\{\{i_1\},\{i_2,i_3,...,
i_m\}\}$ then there exists  $p>0$ and two parties say $P_1$ and
$P_2$ such that,
\begin{equation}
\Psi \storedchem ({\rm EPR}\upp{P_1P_2})^{\otimes p} \enspace.
\end{equation}
} \end{lem} {\bf Proof}: We argue by induction on the number of
parties $m$. The first non-trivial case is when $m=2$. Here
entanglement gambling protocols \cite{BBPS:96} we discussed in the
introduction guarantee the result. So let us assume that the
result is true for $m<k$.  We need to prove that it is true for
$m=k>2$. For this we will use the idea of entanglement of
assistance \cite{DFMSTU:98}. We let $A = i_1$ be the helper and $B
= i_2$ be the first party and $\{i_3,i_4, ..., i_m\}= C$ be the
(composite) second party. Consider the entanglement of assistance
of $\rho\upp{BC}$. If it is zero then from the result on zero
entanglement of assistance from \cite{DFMSTU:98} implies that
either $\rho\upp{BC}=\rho\upp{B}\otimes \proj{\psi\upp{C}}$ or
 $\rho\upp{BC}=\proj{\psi\upp{B}} \otimes \rho\upp{C}$.  Then
either $\Psi=\psi\upp{AB}\otimes\psi\upp{C}$ or
$\Psi=\psi\upp{AC}\otimes\psi\upp{B}$.  In the first case, since
$\Psi$ was entangled across the partition $\{\{i_1\},\{i_2,i_3,
..., i_m\}\}$, $\psi\upp{i_1,i_2}$ has to be entangled, in which
case we apply the $m=2$ case to get an EPR pair between $i_1$ and
$i_2$.  Similarly for the second case $ \psi\upp{i_1,i_3, ...,
i_m}$ must be entangled across the cut
$\{\{i_1\},\{i_3,...,i_m\}\}$,  this by the induction hypothesis
can give an EPR pair between some two parties.  If the
entanglement of assistance is not zero, then A can help B and $C$
to get (with finite probability) an entangled state $\psi\upp{BC}$
i.e. state $\psi\upp{i_2,i_3,...,i_m}$ that is entangled across
the partition $\{\{i_2\},\{i_3,..., i_m\}\}$.  This by the
induction hypothesis can give an EPR pair between some two
parties.  Thus the result is proved.

Note that the result does not require multiple copies of the
starting state.  Note that for proving the above result we used
the necessary and sufficient condition for a state to have zero
entanglement of assistance. It is quite reasonable that the
entanglement of assistance would be useful for a multipartite
scenario, since the motivation for it relies on a three party
scenario.

Now we generalize the stronger version of bipartite entanglement
gambling shown in Eq. \ref{eq:gengamble}. The generalization
involves showing that any irreducible $m$-party state can generate
any other $m$-party state (pure or mixed) with positive
probability using the multicopy stochastic LOCC operations.  We
prove this by showing that we can get an EPR pair between every
pair of parties from any irreducible $m$-partite pure state.  Then
using teleportation, any other state can be generated from these
EPR pairs.  We state this result below.

\begin{theo}
\label{theo:purtoanyEPRB} \label{theo:cat:dis} If state $\Psi$ is
an irreducible $m$-partite state then for any two parties say
$P_2$ and $P_1$ there exists $p>0$ such that,
\begin{equation}
\Psi \mulstoredchem\dnn{LOCC} (\mathrm{EPR}^{P_1P_2})^{\otimes p}
\enspace.
\end{equation}
\end{theo}
{\em Proof}: To prove this we argue by induction on the number of
parties $m$. The first non-trivial case is when $m=2$. Since the
state is irreducible, it is an entangled bipartite state and we
get the result directly from lemma \ref{lem:someEPR}. Assuming the
result to be true for $m<k$, we show that it is true for $m=k$.
Since $\Psi$ is irreducible, by lemma \ref{lem:someEPR} we can
stochastically get an EPR pair between some two parties say $A$
and $B$. If these two are the required parties $P_1$ and $P_2$
then we are done. Otherwise by teleportation through these EPR
pairs, the parties $A$ and $B$ can implement any operation they
could if they were in the same lab. Thus we can look on them as
forming a composite party say $\tilde{A}$. Then we have reduced
the problem to the $m=k-1$ partite case thus proving the result.
\medskip

\section{The Power of a Little Quantum Communication}
In this section we will prove the main result  --- For pure states
asymptotic LOCCq transformations are only as powerful as
asymptotic LOCC state transformations.  First we need to define
these notions of state transformations.

State $\Psi$ is said to be {\em asymptotically LOCC transformable}
state $\Phi$, written as $\Psi \asredchem\dnn{LOCC} \Phi$, if and
only if
\begin{eqnarray}
\forall_{\delta>0,\epsilon>0} \;\exists_{n,n',{\mathcal L}} \;\;
\abs{(n/n')-1}<\delta \;\;{\rm and}& & \nonumber \\
F({\mathcal L}(\Psi^{\otimes n'}),\Phi^{\otimes n}) \ge 1-\epsilon
\enspace & &. \label{defasred}
\end{eqnarray}
Here ${\mathcal L}$ is a multi-locally implementable superoperator
that converts $n'$ copies of $\Psi$ into a high fidelity
approximation to $n$ copies of $\Phi$.  Thus asymptotic
reducibility captures the possibility of state transformations as
the number of source and target copies tends to infinity.  Also
note that if $\psi \mulstoredchem \phi^{\otimes p}$ then $\psi
\asredchem \phi^{\otimes p}$ because of the properties of a
binomial distribution with probability $p$ of success.

Asymptotic reducibilities can have non-integer yields.  This can
be expressed using tensor exponents that take on any nonnegative
real value, so that $\Psi^{\otimes y} \asredchem \Phi^{\otimes x}$
denotes
\begin{eqnarray}
\forall_{\delta>0} \;\exists_{n,n',} \;\;
\abs{(n/n')-x/y}<\delta \;\;{\rm and}& & \nonumber \\
F({\mathcal L}(\Psi^{\otimes n'}) ,\Phi^{\otimes n})  \ge
1-\epsilon \enspace & &.
\end{eqnarray}
In this case we say $x/y$ is the asymptotic efficiency or yield
with which $\Phi$ can be obtained from $\Psi$.  This justifies the
notation used while writing the stochastic state transformations.

A stronger version of asymptotic LOCC state transformation is
obtained if we allow a sublinear amount of quantum communication
during the transformation process.  This is called (asymptotic
LOCCq) state transformation.  We say  state $\Psi$ is {\em
asymptotically LOCCq transformable} to state $\Phi$, written as
$\Psi \asredchem\dnn{LOCCq} \Phi$ if and only if
\begin{eqnarray}
\forall_{\delta>0,\epsilon>0} \;\exists_{n,k,{\mathcal L}} \;\;
(k/n)<\delta \;\;{\rm and}& & \nonumber \\
F({\mathcal L}(\Gamma^{\otimes k}\otimes\Psi^{\otimes n})
,\Phi^{\otimes n})  \ge 1-\epsilon \enspace & &,
\label{defasLOCCq}
\end{eqnarray}
where $\Gamma$ denotes the $m$-Cat state $\ket{0^{\otimes
m}}+\ket{1^{\otimes m}}$. The $m$-Cat states used here are a
convenient way of allowing a sublinear amount $o(n)$ of quantum
communication, since  they can be used as described in
\cite{BPRST:01} to generate EPR pairs between any two parties
which in turn can be used to teleport quantum data between the
parties. The $o(n)$ quantum communication allows the definition to
be simpler in one respect: a single tensor power $n$ can be used
for the input state $\Psi$ and output state $\Phi$, rather than
the separate powers $n$ and $n'$ used in the definition of
ordinary asymptotic LOCC reducibility without quantum
communication, because any $o(n)$ shortfall in number of copies of
the output state can be made up by using the Cat states to
synthesize the extra output states de novo. This definition is
more natural than that for ordinary asymptotic LOCC reducibility
in that the input and output states are allowed to differ in any
way that can be repaired by an $o(n)$ expenditure of quantum
communication, rather than only in the specific way of being $n$
versus $n'$ copies of the desired state where $n-n'$ is $o(n)$.

Clearly $\asredchem\dnn{LOCC}$ implies $\asredchem\dnn{LOCCq}$
because as discussed above asymptotic LOCC state transformation is
a special case of LOCCq state transformations.  An important
question is whether LOCCq state transformations are stronger.  It
turns out that LOCCq state transformations are not stronger than
asymptotic LOCC for pure states.  This constitutes the main result
of the paper.

We start by showing that a state that is factorizable across some
cut can give rise to only states that are factorizable across that
cut under asymptotic LOCCq state transformations.  We prove this
in the following lemma.
\begin{lem}
\label{lem:trivialLOCCq} {\rm : Given state $\Psi$ that is
factorizable across the partition $\{X,\bar{X}\}$ and that
$\Psi\asredchem\dnn{LOCCq}\Phi$, then $\Phi$ must be factorizable
across the same partition. }\end{lem} {\bf Proof}: This is
essentially a two party problem, with $X$ and $\bar{X}$ as the two
compound  parties. We argue by contradiction. Suppose $\Phi$ was
non-factorizable across the partition $\{X, \bar{X}\}$ with
bipartite entanglement $x>0$. Then $n$ copies of $\Phi$ would have
$nx$ bipartite entanglement across the partition. However, since
$\Psi$ has no entanglement across the partition and since LOCCq
protocols only allow a sub linear amount of $m$-Cat states along
with LOCC, they cannot increase the entanglement across the
partition by more than $o(n)$. Thus, no asymptotic LOCCq protocol
can give rise to $\Phi$ starting from $\Psi$.
\medskip

Now we prove that for irreducible pure states, asymptotic LOCCq
and asymptotic LOCC are equally powerful.
\begin{lem}
\label{lem:irrLOCCqeqLOCC} For an irreducible $m$-partite pure
state $\Psi$ and any arbitrary state $\Phi$,
\begin{equation}
\Psi \asredchem\dnn{LOCCq} \Phi \Leftrightarrow \Psi
\asredchem\dnn{LOCC} \Phi \enspace.
\end{equation}
\end{lem}
\noindent{\em Proof} Since $\Psi$ is irreducible, it is cat
distillable from theorem \ref{theo:cat:dis}. Hence we can use a
$o(n)$ copies of $\Psi$ to generate $o(n)$ copies of $m$Cat by
LOCC, which we can use for the $o(n)$ quantum communication
required for LOCCq. Since only $o(n)$ extra copies of $\Psi$ are
required than the LOCCq protocol, this does not change the yield
asymptotically, and hence the LOCCq protocol can be simulated by
an LOCC protocol.  This proves the result.

Now we are ready to combine the results from the above lemmas to
prove the general result as the theorem below.
\begin{theo}
\label{theo:LOCCqeqLOCC} For $m$-partite pure states $\Psi$ and
$\Phi$,
\begin{equation}
\Psi \asredchem\dnn{LOCCq} \Phi \Leftrightarrow \Psi
\asredchem\dnn{LOCC} \Phi \enspace.
\end{equation}
\end{theo}
\noindent{\em Proof} We argue by induction on the number of
parties $m$. Consider the first non-trivial case $m=2$. If $\Psi$
is irreducible, then  theorem \ref{theo:purtoanyEPRB} along with
lemma \ref{lem:irrLOCCqeqLOCC} gives us the result. If $\Psi$ is
factorizable, in this case  a product state, then by lemma
\ref{lem:trivialLOCCq} $\Phi$ must be a product state too and thus
can be created trivially by LOCC operations. Now let the theorem
be true for all $m< k$, then we show that it is true for $m=k$. If
$\Psi$ is irreducible, then theorem \ref{theo:purtoanyEPRB}
 along with lemma \ref{lem:irrLOCCqeqLOCC} gives us the result. Otherwise
$\Psi$ is factorizable across some cut $\{X,\bar{X}\}$.  Then
lemma \ref{lem:trivialLOCCq} implies that $\Phi$ is factorizable
across the same cut i.e., $\Phi = \phi_1\upp{X} \otimes
\phi_2\upp{\bar{X}}$. Applying this theorem for $m<k$, to the
states $\phi_1\upp{X}$ and $\phi_2\upp{X}$ we have the result.
\medskip

Thus we have shown that LOCC and LOCCq are equivalent for pure
states.

Let us now turn our attention to an application of entanglement
gambling to multipartite distillability.



\section{Entanglement Gambling and Multipartite Distillability}
In this section we will study some implications of the
entanglement gambling result to the notion of distillability in
multi-party systems.

One of the main problems with defining distillable entanglement
for multiple parties is that since there are many different kinds
of entanglement, it is impossible to maximize over the yield of
all those states.  However, we may easily generalize the notion of
distillability from the bipartite scenario to get the following
general definition of distillability: {\em We say $\rho$ is
distillable if and only if} $\rho \asredchem \Psi^{\otimes x}$ for
some positive $x$, where $\psi$ is some entangled pure state.

 However, operationally it is more useful to have EPR pairs or
Cat-states as the target state to be produced in the distillation
procedure, since they can then directly be used to achieve other
information processing tasks. Thus, one may define EPR
distillability as: {\em We say $\rho$ is distillable if and only
if} $\rho \asredchem \Psi^{\otimes x}$ for some positive $x$,
where $\Psi$ is an EPR pair between some pair of parties.
Similarly, one may define Cat distillability as EPR
distillability, except the target state $\Psi$ is now required to
be a $m$-Cat state.

The relation between general distillability and EPR/Cat
distillability is an interesting issue.  In the bipartite case
since any pure entangled state can be converted to an EPR pair, it
turns out that EPR-distillability and distillability are
identical. Clearly we would want this property to be true for
multipartite states also.  Clearly all we need to show is that any
entangled multipartite pure state can give some EPR pair
asymptotically, since then the entangled pure state $\Psi$ in the
general definition distillability above, can be converted to an
EPR pair.  This is precisely the result of lemma
\ref{theo:purtoanyEPRB}!  Thus we can say that {\em A $m$-partite
state $\rho$ is distillable if and only if it is EPR distillable}.

Clearly, if a state is Cat-distillable it is also distillable and
EPR distillable.  Clearly, the converse is not true in general.
Cat-distillable states are interesting because they can generate
all other states and hence form a minimal entanglement generating
set (MEGS), that is a minimal set of states that can generate any
other state under asymptotic LOCC.  Since the reversibility of the
state transformations is not required, this is a very coarse
grained entanglement measure.   Let us consider a state that is
factorizable across some cut of parties $\{{X},{\bar{X}}\}$. Then
it cannot be cat-distillable because that would imply that a
separable bipartite state can be made into an entangled one with
LOCC operations, which we know is impossible. Thus only
irreducible states can be cat-distillable.  Then lemma
\ref{theo:purtoanyEPRB} shows that any irreducible  pure state is
cat-distillable. Putting these together we see that {\em a pure
state is cat-distillable if and only if it is irreducible}. But
dropping the requirement of reversibility still gives a
qualitative broad picture of multipartite entanglement. This is
analogous to classifying bipartite mixed states as distillable and
undistillable to get a coarse grained measure of distillable
entanglement. In this light, the result is very satisfying because
it says that: If we allow ourselves to waste entanglement during
transformation of states, then any irreducible state is equivalent
to any other, and is more powerful entanglement-wise than any
factorizable state, thus giving a hierarchy of qualitatively
different entangled states which factorize into irreducible parts
of various sizes.

A natural question is whether a non-factorizable mixed state is
also cat-distillable. This obviously is false, because that would
imply separable but non-factorizable bipartite states could
generate entanglement, which we know cannot happen. So we need to
generalize the idea of irreducibility to mixed states.  The
natural way to do this is by replacing the idea of factorizability
to that of separability. So we say that $\Psi$ is reducible across
a partition $\{X,\bar{X}\}$ of parties if it is separable across
that partition. We say a state is irreducible if it is not
separable across any partition of the parties.  This
generalization is not useful because of the existence of bound
entangled states, that is states which are inseparable but not
distillable.  So, we could generalize irreducibility to mixed
states using distillability across cuts: We say a state is
irreducible if it is distillable across all cuts.  Given this
generalization of the definition, it is an open question whether
cat-distillability and irreducibility are equivalent for mixed
states, because lemma \ref{lem:someEPR} does not hold for mixed
states in general \cite{S:00, SST:00}.  This just means that our
approach from theorem \ref{theo:cat:dis} won't carry over to mixed
states.


\section{Discussions and Conclusions}
In this paper we have shown that asymptotic LOCC and LOCCq state
transformations are equally powerful for pure states.  Clearly an
important question is whether LOCCq is more powerful than
asymptotic LOCC for mixed states.  Obviously, for cat-distillable
(mixed) states our result showing that the two have equal power
should hold since we can use $o(n)$ Cat-states to achieve $o(n)$
quantum communication. Thus, the open question is mainly regarding
the mixed states that are not cat-distillable. This is an
important future direction. One possible way to get the full mixed
state result just as we did for pure states, using induction and
showing that factorizable states can only give rise to
factorizable states under LOCCq transformations, leads to the
problem of how to define irreducible mixed states such that they
are cat-distillable and at the same time would facilitate an
inductive argument.

We have shown here that any irreducible (non-factorizable) pure
state is cat-distillable, however our protocols are not very
efficient, and that was not the goal either.  However, in reality,
we need cat-distillable protocols that are efficient.  Finding
such protocols is another important future direction.

\begin{acknowledgments}
AVT acknowledges support from Defense Advanced Research Projects
Agency (DARPA) and the Air Force Laboratory, Air Force Material
Command, USAF, under Contract No. F30602-01-2-0524, also from the
USA Army Research Office, under grants DAAG-55-98-C-0041, and
DAAG-55-98-1-0366, and support from IBM Research.  JAS
acknowledges support from the USA Army Research Office, under
grant DAAG-55-98-C-0041.

\end{acknowledgments}

\bibliography{qit}

\begin{thebibliography}{12}
\expandafter\ifx\csname natexlab\endcsname\relax\def\natexlab#1{#1}\fi
\expandafter\ifx\csname bibnamefont\endcsname\relax
  \def\bibnamefont#1{#1}\fi
\expandafter\ifx\csname bibfnamefont\endcsname\relax
  \def\bibfnamefont#1{#1}\fi
\expandafter\ifx\csname citenamefont\endcsname\relax
  \def\citenamefont#1{#1}\fi
\expandafter\ifx\csname url\endcsname\relax
  \def\url#1{\texttt{#1}}\fi
\expandafter\ifx\csname urlprefix\endcsname\relax\def\urlprefix{URL }\fi
\providecommand{\bibinfo}[2]{#2}
\providecommand{\eprint}[2][]{\url{#2}}

\bibitem[{\citenamefont{Bennett et~al.}(2001)\citenamefont{Bennett, Popescu,
  Rohrlich, Smolin, and Thapliyal}}]{BPRST:01}
\bibinfo{author}{\bibfnamefont{C.~H.} \bibnamefont{Bennett}},
  \bibinfo{author}{\bibfnamefont{S.}~\bibnamefont{Popescu}},
  \bibinfo{author}{\bibfnamefont{D.}~\bibnamefont{Rohrlich}},
  \bibinfo{author}{\bibfnamefont{J.~A.} \bibnamefont{Smolin}},
  \bibnamefont{and} \bibinfo{author}{\bibfnamefont{A.~V.}
  \bibnamefont{Thapliyal}}, \bibinfo{journal}{Phys.\ Rev. A}
  \textbf{\bibinfo{volume}{63}}, \bibinfo{pages}{012307}
  (\bibinfo{year}{2001}).

\bibitem[{\citenamefont{Bennett et~al.}(1993)\citenamefont{Bennett, Brassard,
  Crépeau, Jozsa, Peres, and Wootters}}]{BBCJPW:93}
\bibinfo{author}{\bibfnamefont{C.~H.} \bibnamefont{Bennett}},
  \bibinfo{author}{\bibfnamefont{G.}~\bibnamefont{Brassard}},
  \bibinfo{author}{\bibfnamefont{C.}~\bibnamefont{Crépeau}},
  \bibinfo{author}{\bibfnamefont{R.}~\bibnamefont{Jozsa}},
  \bibinfo{author}{\bibfnamefont{A.}~\bibnamefont{Peres}}, \bibnamefont{and}
  \bibinfo{author}{\bibfnamefont{W.~K.} \bibnamefont{Wootters}},
  \bibinfo{journal}{Phys.\ Rev.\ Lett.} \textbf{\bibinfo{volume}{70}},
  \bibinfo{pages}{1895} (\bibinfo{year}{1993}).

\bibitem[{\citenamefont{Bennett and Wiesner}(1992)}]{BW:92}
\bibinfo{author}{\bibfnamefont{C.~H.} \bibnamefont{Bennett}} \bibnamefont{and}
  \bibinfo{author}{\bibfnamefont{S.~J.} \bibnamefont{Wiesner}},
  \bibinfo{journal}{Phys.\ Rev.\ Lett.} \textbf{\bibinfo{volume}{69}},
  \bibinfo{pages}{2881} (\bibinfo{year}{1992}).

\bibitem[{\citenamefont{Bennett et~al.}(1999)\citenamefont{Bennett, Shor,
  Smolin, and Thapliyal}}]{BSST:99}
\bibinfo{author}{\bibfnamefont{C.~H.} \bibnamefont{Bennett}},
  \bibinfo{author}{\bibfnamefont{P.~W.} \bibnamefont{Shor}},
  \bibinfo{author}{\bibfnamefont{J.~A.} \bibnamefont{Smolin}},
  \bibnamefont{and} \bibinfo{author}{\bibfnamefont{A.~V.}
  \bibnamefont{Thapliyal}}, \bibinfo{journal}{Phys.\ Rev.\ Lett.}
  \textbf{\bibinfo{volume}{83}}, \bibinfo{pages}{3081} (\bibinfo{year}{1999}).

\bibitem[{\citenamefont{Bennett et~al.}(2002)\citenamefont{Bennett, Shor,
  J.A.Smolin, and A.V.Thapliyal}}]{BSST:02}
\bibinfo{author}{\bibfnamefont{C.~H.} \bibnamefont{Bennett}},
  \bibinfo{author}{\bibfnamefont{P.}~\bibnamefont{Shor}},
  \bibinfo{author}{\bibnamefont{J.A.Smolin}}, \bibnamefont{and}
  \bibinfo{author}{\bibnamefont{A.V.Thapliyal}}, \bibinfo{journal}{IEEE
  Transactions on Information Theory} \textbf{\bibinfo{volume}{48(10)}},
  \bibinfo{pages}{2637} (\bibinfo{year}{2002}).

\bibitem[{\citenamefont{Shor}(1994)}]{S:94}
\bibinfo{author}{\bibfnamefont{P.~W.} \bibnamefont{Shor}}, in
  \emph{\bibinfo{booktitle}{Proc. 35nd Annual Symposium on Foundations of
  Computer Science}}, edited by
  \bibinfo{editor}{\bibfnamefont{S.}~\bibnamefont{Goldwasser}}
  (\bibinfo{publisher}{IEEE Computer Society Press}, \bibinfo{year}{1994}), pp.
  \bibinfo{pages}{124--134}.

\bibitem[{\citenamefont{Bennett and Brassard}(December 1984)}]{BB:84}
\bibinfo{author}{\bibfnamefont{C.}~\bibnamefont{Bennett}} \bibnamefont{and}
  \bibinfo{author}{\bibfnamefont{G.}~\bibnamefont{Brassard}}, in
  \emph{\bibinfo{booktitle}{Proceedings of IEEE International Conference on
  Computers Systems and Signal Processing, Bangalore India}}
  (\bibinfo{year}{December 1984}), pp. \bibinfo{pages}{175--179}.

\bibitem[{\citenamefont{Bennett et~al.}(1996)\citenamefont{Bennett, Bernstein,
  Popescu, and Schumacher}}]{BBPS:96}
\bibinfo{author}{\bibfnamefont{C.~H.} \bibnamefont{Bennett}},
  \bibinfo{author}{\bibfnamefont{H.~J.} \bibnamefont{Bernstein}},
  \bibinfo{author}{\bibfnamefont{S.}~\bibnamefont{Popescu}}, \bibnamefont{and}
  \bibinfo{author}{\bibfnamefont{B.}~\bibnamefont{Schumacher}},
  \bibinfo{journal}{Phys.\ Rev. A} \textbf{\bibinfo{volume}{53}},
  \bibinfo{pages}{2046} (\bibinfo{year}{1996}).

\bibitem[{\citenamefont{D$\mathrm{\ddot{u}}$r}(2001)}]{D:01}
\bibinfo{author}{\bibfnamefont{W.}~\bibnamefont{D$\mathrm{\ddot{u}}$r}},
  \bibinfo{journal}{Phys.\ Rev. Lett.} \textbf{\bibinfo{volume}{87}},
  \bibinfo{pages}{230402} (\bibinfo{year}{2001}).

\bibitem[{\citenamefont{DiVincenzo et~al.}(1999)\citenamefont{DiVincenzo,
  Fuchs, Smolin, Thapliyal, and Uhlmann}}]{DFMSTU:98}
\bibinfo{author}{\bibfnamefont{D.~P.} \bibnamefont{DiVincenzo}},
  \bibinfo{author}{\bibfnamefont{C.~A.} \bibnamefont{Fuchs}},
  \bibinfo{author}{\bibfnamefont{J.~A.} \bibnamefont{Smolin}},
  \bibinfo{author}{\bibfnamefont{A.}~\bibnamefont{Thapliyal}},
  \bibnamefont{and} \bibinfo{author}{\bibfnamefont{A.}~\bibnamefont{Uhlmann}},
  in \emph{\bibinfo{booktitle}{Proceedings of the First NASA International
  Conference on Quantum Computing and Quantum Communications, 17-20th February
  1998, Palm Springs, CA}}, edited by \bibinfo{editor}{\bibfnamefont{C.~P.}
  \bibnamefont{Williams}} (\bibinfo{publisher}{Springer-Verlag, Heidelberg,
  Germany}, \bibinfo{year}{1999}), vol. \bibinfo{volume}{1509} of
  \emph{\bibinfo{series}{Lecture Notes in Computer Science}}.

\bibitem[{\citenamefont{Smolin}(2001)}]{S:00}
\bibinfo{author}{\bibfnamefont{J.~A.} \bibnamefont{Smolin}},
  \bibinfo{journal}{Phys. Rev. A} \textbf{\bibinfo{volume}{63}},
  \bibinfo{pages}{032306} (\bibinfo{year}{2001}).

\bibitem[{\citenamefont{Shor et~al.}()\citenamefont{Shor, Smolin, and
  Thapliyal}}]{SST:00}
\bibinfo{author}{\bibfnamefont{P.~W.} \bibnamefont{Shor}},
  \bibinfo{author}{\bibfnamefont{J.~A.} \bibnamefont{Smolin}},
  \bibnamefont{and} \bibinfo{author}{\bibfnamefont{A.~V.}
  \bibnamefont{Thapliyal}}, \bibinfo{journal}{submitted to Phys.\ Rev. Lett.}
  (????).

\end{thebibliography}

\end{document}